%% file: ase2010.tex
\documentclass{sig-alternate}
\usepackage{listings}
\newcommand{\comment}[1]{}

\begin{document}

\lstset{language=C,basicstyle=\small}
\lstset{numbers=left, numberstyle=\tiny, stepnumber=1, numbersep=5pt}
\lstset{tabsize=2}
\lstset{firstnumber=1}
\lstset{frame=single}
\lstset{
  language={C},
  morekeywords={assert,uchar}
}

%

\title{Bounded Model Checking of Multi-threaded Software using SMT solvers}
%
%
%
%
%

\numberofauthors{2} 
%
\author{
%
%
\alignauthor
Lucas Cordeiro\\
       \affaddr{University of Southampton}\\
       \email{lcc08r@ecs.soton.ac.uk}        
\alignauthor
Bernd Fischer\\
       \affaddr{University of Southampton}\\
       \email{b.fischer@ecs.soton.ac.uk}              
}

\maketitle
\begin{abstract}
The transition from single-core to multi-core processors has made multi-threaded software an important subject in computer aided verification. Here, we describe and evaluate an extension of the ESBMC model checker to support the verification of multi-threaded software with shared variables and locks using bounded model checking (BMC) based on Satisfiability Modulo Theories (SMT). We describe three approaches to model check multi-threaded software and our modelling of the synchronization primitives of the Pthread library. In the lazy approach, we generate all possible interleavings and call the BMC procedure on each of them individually, until we either find a bug, or have systematically explored all interleavings. In the schedule recording approach, we encode all possible interleavings into one single formula and then exploit the high speed of the SMT solvers. In the underapproximation-widening approach, we reduce the state space by abstracting the number of state variables and interleavings from the proofs of unsatisfiability generated by the SMT solvers. In all three approaches, we use partial-order reduction (POR) techniques to reduce the number of interleavings explored. Experiments show that our approaches can analyze larger problems and substantially reduce the verification time compared to state-of-the-art techniques that combine classic POR methods with symbolic algorithms and others that implement the Counter-Example Guided Abstraction Refinement technique.
\end{abstract}



\input{01-introduction}


\input{02-bmc-multi-threaded-programs}

\input{03-synchronization-primitives}

\input{04-experimental-evaluation}

\input{05-related-work}

\input{06-conclusion}



{\small\noindent{\bf Acknowledgments.} 
We thank D. Kroening for the fruitful discussions about the starting point of this work and J. Rathke for many helpful discussions about multi-process systems. We also thank S. Yiu for his work in his MSc Thesis~\cite{yiu10} on model checking concurrent programs with SAT. We thank J. Colley, D. Nicole, R. Silva, and R. Quigley for their comments on a draft version.
}

%

\end{document}

%% file: 01-introduction.tex
\section{Introduction}
\label{01-introduction}

Embedded computer systems are used in a wide range of sophisticated applications, such as mobile phones or set-top boxes providing internet connectivity. The functionality demanded in such applications has increased significantly and an increasing number of functions are implemented in software rather than hardware.\ Thus, multi-core processors with scalable shared memory have become popular in embedded systems. In turn, the verification of the software design and the correctness of its multi-threaded implementations has become increasingly difficult. 

Bounded model checking (BMC) has already been successfully applied to verify embedded software and discover subtle errors in real designs~\cite{handbook09}.\ BMC generates verification conditions (VCs) that reflect the exact path in which a statement is executed, the context in which a given function is called, and the bit-accurate representation of the expressions.\ Proving the validity of these VCs remains a major performance bottleneck in verifying embedded software, despite attempts to cope with increasing system complexity by applying SMT (Satisfiability Modulo Theories)~\cite{Armando09,Cordeiro09,Ganai06}. 

Recently, there have been attempts to extend BMC to the verification of multi-threaded software~\cite{GanaiG08,KahlonSG09,KahlonWG09,RabinovitzG05}.\ The main challenge is the state space explosion problem, as the number of interleavings grows exponentially with the number of threads and program statements.\ However, two important observations help us: \textit{(i)} SMT-based BMC finds counter-examples very quickly~\cite{Cordeiro09} and \textit{(ii)} SMT solvers produce unsatisfiable cores that allow us to remove logic that is not relevant to a given property~\cite{McMillanA03}. Grumberg et~al.\ \cite{GrumbergLST05} realized that the unsatisfiable cores generated by the solvers can also be used to control the number of allowed interleavings of the given set of processes.\ They propose an algorithmic method based on Boolean Satisfiability (SAT) and BMC to model check a multi-process system based on a series of under-approximated models. However, this method does not combine classic partial-order reduction (POR) methods with symbolic algorithms, which limits its usefulness for analyzing and verifying multi-threaded software. It has also not been applied in conjunction with SMT solvers.

In our prior work~\cite{Cordeiro09}, we extended the encodings from previous SMT-based bounded model checkers~\cite{Armando09,Ganai06} to provide more accurate support for variables of finite bit width, bit-vector operations, arrays, structures, unions and pointers.\ Here, we develop three approaches to tackle complexity problems in model checking multi-threaded C software.\ In the lazy approach, we extend the BMC procedure of single-threaded software to multi-threaded software by wrapping it inside a straightforward generate-and-test loop, which generates all possible interleavings and calls the BMC procedure on each of them individually. We stop this loop either when we find a bug, or have systematically explored all interleavings. In the scheduling recording approach, we explore systematically the control-flow graph (CFG) of each thread and encode all the possible execution paths into one single formula, which is then fed into the back-end SMT solver. In our third approach, we extend the under-approximation and widening (UW) algorithm proposed in~\cite{GrumbergLST05} with the purpose of addressing the verification of real-world C code using different background theories and SMT solvers.\ 

We also implement partial order reduction algorithms~\cite{AlurBHQR01} in our three approaches and propose a comprehensive SMT-based BMC procedure to support the checking of multi-threaded programs that utilize the synchronization primitives of the POSIX Pthread Library~\cite{Mueller93alibrary}. To our knowledge, this work marks the first application of the UW algorithm combined with POR techniques to model check non-trivial multi-threaded C software. Experiments obtained with ESBMC show that our approaches can analyze larger problems and substantially reduce the verification time compared to state-of-the-art techniques that combine classic POR methods with symbolic algorithms and others that implement the Counter-Example Guided Abstraction Refinement (CEGAR) technique.

%% file: 02-bmc-multi-threaded-programs.tex
\section{Bounded Model Checking of \\* Multi-threaded Software}
\label{03-proposed-method}

In BMC, the program to be analyzed is modelled as a state transition system, which is built by extracting its behaviour from the CFG. This graph is used as part of a translation process from program text to single static assignment (SSA) form. Each thread is modelled as a CFG where nodes represent program statements and edges represent transitions. A state transition system $M=(S, R, S_0)$ is an abstract machine that consists of a set of states $S$ (where $S_{0} \subseteq S$ represents the set of initial states) and transitions $R$ between states, i.e., for each $\gamma \in R$, $\gamma \subseteq S \times S$. A state $s \in S$ consists of the value of the program counter \emph{pc} and the values of all program variables.\ An initial state $s_{0}$ assigns the initial program location of the CFG to the \emph{pc}. We identify each transition $\gamma=(s_i,s_{i+1})$ between two states $s_{i}$ and $s_{i+1}$ with a logical formula $\gamma(s_i,s_{i+1})$ that captures the constraints on the values of the program counter and the program variables. 

\begin{figure}[ht]
\centering
\includegraphics[scale=0.3]{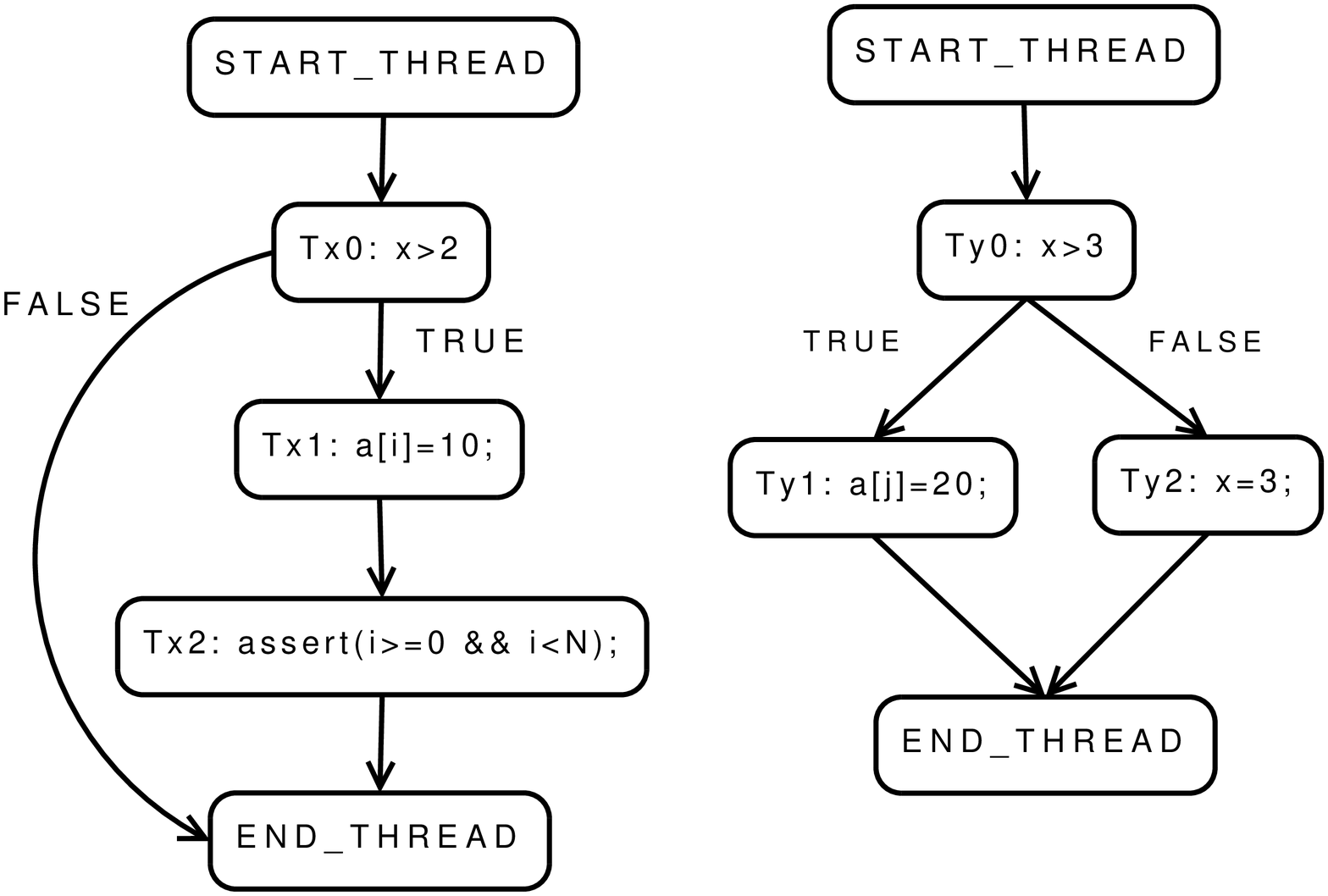}
\caption{Control-flow graph of two threads.}
\label{figure:control-point}
\end{figure}

As a running example, we consider the C program in Figure~\ref{figure:A-multi-threaded-C-program}, which consists of two threads that are created using the Pthread library~\cite{Mueller93alibrary}. Note that our example contains a subtle bug (array lower bound) in line 9, because function $nondet\_uint()$ might return non-deterministically a negative integer number and as a result the {\tt assert} macro (line 10) fails. Figure~\ref{figure:control-point} shows the CFG representation of the two threads $T_{X}$ and $T_{Y}$. After creation, they are at the control points $T_{X_{0}}$ and $T_{Y_{0}}$ respectively, and since $x==2$ (see line 3 of Figure~\ref{figure:A-multi-threaded-C-program}), both tests $x > 2$ and $x > 3$ are false. If we schedule $T_{X}$ first, it will not be enabled, and we can transition to the next state only by switching to $T_{Y}$ and executing only the program statement $Y_{1}$ (i.e., $x=3$, see line 18) before terminating. If we continue exploring the remaining interleavings, we schedule $T_{Y}$ first, and the execution of $Y_{1}$ makes the test $x>2$ in line 7 true, thus enabling $T_{X}$ to progress and transition through $X_{0}$ and $X_{1}$, i.e., we execute program statements $x=3$, $a[i]=*((int *)arg)$, and  $assert(i>=0 \: \&\& \: i<N)$). Note that, as in \cite{GanaiG08} we do not model context switches inside the execution of individual statements, to avoid exploring additional interleavings. This approach is safe as long as statements only read or write a single global variable, but is an under-approximation to programs that contain statements involving multiple global variables. However, with the benchmarks that are publicly available, we have not encountered any problems in practice.

\begin{figure}[ht]
\centering
\begin{lstlisting}
#include <pthread.h>
#define N 10
int a[N], i, j=1, x=2;
int nondet_uint();
void *Tx(void *arg)
{
  if (x>2)
  {
    a[i]=*((int *)arg);   //X0
    assert(i>=0 && i<N);  //X1
  }
}
void *Ty(void *arg)
{
  if (x>3)
    a[j]=*((int *)arg);  //Y0
  else
    x=3;  							 //Y1
}
int main()
{
  pthread_t id1, id2; 
  int arg1=10, arg2=20;
  i=nondet_uint();  
  pthread_create(&id1, NULL, Tx, &arg1);
  pthread_create(&id2, NULL, Ty, &arg2);
}
\end{lstlisting}
\caption{A multi-threaded C program with violated property.}
\label{figure:A-multi-threaded-C-program}
\end{figure}

Formally, given a transition system \textit{M}, a property $\phi$, and a bound \textit{k}, BMC unrolls the system \textit{k} times and translates it into a verification condition $\psi$ such that $\psi$ is satisfiable \emph{iff} $\phi$ has a counterexample of depth less than or equal to \textit{k}.\ The model checking problem associated with SMT-based BMC for checking linear-time temporal logic (LTL) properties is then formulated by constructing the logical formula:

\begin{equation}
\label{bounded-model-checking}
\psi^{k} = \overbrace{I\left(s_{0}\right) \wedge \bigwedge^{n}_{j=1} \bigwedge^{k-1}_{i=0} \gamma_{j} \left(s_{i},s_{i+1}\right)}^{constraints} \wedge \overbrace{\bigwedge^{n}_{j=1} P_{j}\left(s_{k}\right)}^{property}
\end{equation}

\noindent where $P_{j}\left(s_{k}\right)$ represents a LTL property $\phi$ in step $k$ of thread $j$, $I$ is the function for the set of initial states of $M$ and $\gamma_{j} \left(s_{i},s_{i+1}\right)$ is the function of the transition relation of thread $j$ at time steps $i$ and $i+1$. Hence, the formula $\bigwedge^{n}_{j=1} \bigwedge^{k-1}_{i=0} \gamma_{j} \left(s_{i},s_{i+1}\right)$ represents the set of all executions of $n$ threads up to the length $k$ or less. $P_{j}\left(s_{k}\right)$ is derived from the property being checked and represents the condition that it is violated by a bounded execution of thread $j$ of length $k$ or less. Note that formula (\ref{bounded-model-checking}) encodes all allowed interleavings of the given threads. 

\subsection{Lazy Approach}
\label{sec:LazyApproach}

Conceptually, the simplest way to extend a bounded model checker for single-threaded software to the multi-threaded case is to wrap it inside a straightforward generate-and-test loop: we just need to generate all possible interleavings and call the BMC procedure on each of them individually, until we either find an error, or have systematically explored all interleavings.

\begin{figure}[ht]
\centering
\includegraphics[scale=0.33]{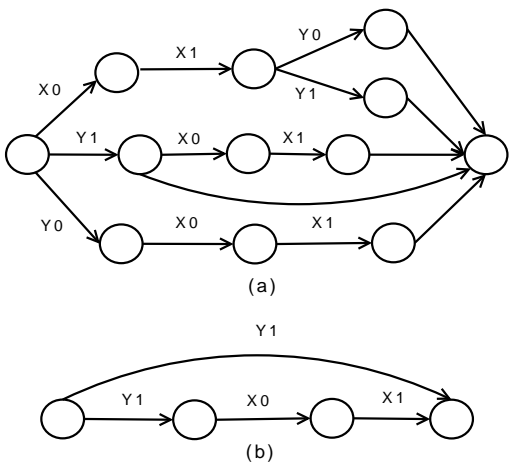}
\caption{(a) All possible thread interleavings in Figure~\ref{figure:control-point} (b) The actual thread interleavings after using information from the front-end.}
\label{figure:all-interleavings}
\end{figure}

On the face of it, this seems to be naive: the number of interleavings can grow very quickly (see Figure~\ref{figure:all-interleavings}(a) for
all possible interleavings in the running example), and we need to invoke the model checker several times, which might slow down the verification
process. 

However, there are several observations that make this approach worthwhile. First we can obviously generate each interleaving, model check it, and stop the generation when we find the first error. In practice, if the program contains any errors, they will be exhibited in a substantial fraction of the interleavings, if not all (experience of \cite{QadeerR05} for real applications), so that we only need to explore a small part of the search space. Second, we obviously do not need to generate the source code of all possible interleavings. Instead, we keep in memory the nodes of all unexplored execution paths and expand them one path at a time. We then construct the VCs for the chosen execution path according to formula (\ref{bounded-model-checking}) and feed it into the SMT solver to check for satisfiability. Third, and most important, we can use information from the front-end to reduce both the number of interleavings to be explored and the size of the formulas sent to the SMT solver. In particular, during the symbolic execution we exploit which transitions are enabled in a given state to drive the exploration of the interleavings. 

In our running example, the transitions from $T_{X_{0}}$ to $T_{X_{1}}$ and from $T_{Y{_0}}$ to $T_{Y_{1}}$ are disabled because we initially have $x=2$. This rules out all interleavings that start with either $X_0$ or $Y_0$ and only leaves those that bypass $T_X$ entirely, or start with $Y_{1}$. Assuming that we explore the thread $T_{X}$ first, in the first iteration we thus build the VCs only for the program statement $Y_{1}$. We then pass the formula (\ref{bounded-model-checking}) to the SMT solver and check its satisfiability. If it is satisfiable, we have found a property violation and we can stop the process. Here, however, it is not satisfiable and we continue to the next iteration by selecting an unexplored path. In the second iteration, we explore $T_{Y}$ first, and select program statement $Y_{1}$, and after that we explore $T_{X}$ and select program statements $X_{0}$ and $X_{1}$. Again, we pass the corresponding version of formula (\ref{bounded-model-checking}) to the SMT solver. Since this is now satisfiable, we can stop the exploration of the execution paths.

In summary, we guide the symbolic execution between the threads and systematically explore all the possible execution paths in a lazy way. This approach can find bugs fast, but as the front-end might invoke the SMT solver repeatedly, once for each possible execution path, it can suffer performance degradation, in particular for correct programs where we need to explore all possible interleavings. The invocation procedure itself is slow and the formula needs to be passed from front-end to back-end several times. Moreover, execution paths that share the same program statements will be unnecessarily checked several times. However, as each formula corresponds to one possible path only, its size is relatively small compared to the schedule recording approach described in the next section and can thus be handled easily by the SMT solver without requiring too much memory.

\subsection{Schedule Recording Approach}
\label{sec:EagerEncoding}

State-of-the-art SMT solvers are built on top of SAT solvers to speed up the performance by exploiting the support for ``conflict clauses'' and non-chronological backtracking~\cite{SilvaS96}. In the schedule recording approach we leverage this and avoid invoking the SMT solver multiple times. We use the symbolic execution engine as before to systematically explore the interleavings, but now we add schedule guards to record in which order the scheduler has executed the program. We then encode all execution paths into one formula, which is finally fed into the SMT solver. However, the number of threads and context switches can grow very large quickly, and easily ``blow-up'' the solver. Given this, there is a clear trade-off between usage of time and memory resources to model check multi-threaded software.

\begin{figure}[ht]
\centering
\includegraphics[scale=0.32]{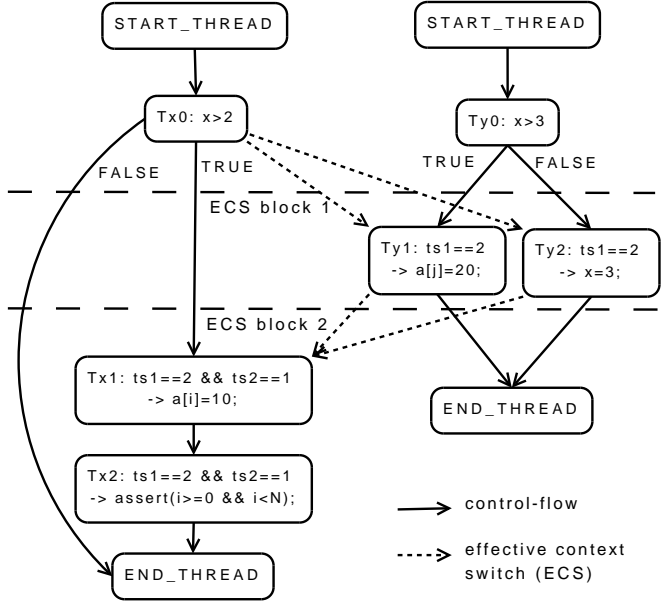}
\caption{Schedule encoding of the example in Figure~\ref{figure:control-point}.}
\label{figure:eager-approach}
\end{figure}

Figure~\ref{figure:eager-approach} illustrates our schedule recording encoding applied to the example in Figure~\ref{figure:control-point}. Since control-flow tests cannot influence the state (as the front-end hoists side-effects out of the tests), we only need to add guards to \textit{effective statements}, i.e., assignments and assertions. Similarly, we only need to record \textit{effective context switches} (ECS), i.e., context switches to an effective statement. These are shown as dashed arrows in Figure~\ref{figure:eager-approach}. Finally, we define an \textit{ECS block} as a sequence of program statements that are executed with no intervening ECS, and give each block a number. Each effective program statement is then prefixed by a schedule guard $ts_{i}=j$ where $i$ is the ECS block number and $j$ is the thread identifier. Its intuitive interpretation is that the guarded statement can only be executed if thread $j$ is scheduled in the $i$-th ECS block. The value of $ts_{i}$ is set by the SMT solver, and determines the order in which the program statements are executed. For example, the guard at $T_{Y2}$ thus encodes that $Y_{1}$ can only be executed if $T_{Y}$ runs in the first ECS block. Note that schedule guards are only necessary but not sufficient conditions for the execution of a statement. For example, $T_{Y1}$ has the same guard as $T_{Y2}$, but $Y_{0}$ cannot be executed using any viable schedule. The guards can also be combined conjunctively and disjunctively to encode more involved schedules. For example, the guard of both $T_{X1}$ and $T_{X2}$ corresponds to a schedule in which $T_{Y}$ ran before switching to $T_{X}$.

The schedule guards are added by the front-end when program statements are executed symbolically and become part of the produced verification conditions. The thread selection variable is a free variable that the SMT solver will try to instantiate with all possible concrete values. The thread number value is a constant that corresponds to the thread identifier. As an example, if the SMT solver chooses $ts_{1}=2$ and $ts_{2}=1$, then the program statement $X_{0}, X_{1}, Y_{0}$, and $Y_{1}$ are all enabled in principle, but which ones are executed depends on the values of the control-flow tests $x>2$ and $x>3$. Note that the ordering of statements within a thread is of course still ensured by the program order semantics, so that $X_{1}$ will not be executed before $X_{0}$. Consequently, all the combinations of the thread selection variables will produce only two different interleavings as follows: $\left\{Y_{1}\right\}$ and $\left\{Y_{1}, X_{0}, X_{1}\right\}$ (cf. Figure~\ref{figure:all-interleavings}(b)). 

Given this, we can define a schedule $SCH$ to determine which interleavings will be considered and encode the guards in formula (\ref{bounded-model-checking}) as:

\begin{equation}
\label{bounded-model-checking-eager}
\psi^{k} = \overbrace{I\left(s_{0}\right) \wedge \bigwedge^{n}_{j=1} \bigwedge^{k-1}_{i=0} \gamma'_{j} \left(s_{i},s_{i+1}\right)}^{constraints} \wedge \overbrace{\bigwedge^{n}_{j=1} P_{j} \left(s_{k}\right)}^{property} \wedge \overbrace{\bigwedge^{k-1}_{i=0} SCH\left(s_{i}\right)}^{schedule}
\end{equation}

\noindent where $\gamma'_{j}$ represents the modified transition relation incorporating the schedule guards added by the front-end and $SCH\left(s_{i}\right)$ represents a constraint on the schedule. If we do not add any constraints, then $\bigwedge^{k-1}_{i=0} SCH\left(s_{i}\right) = true$ and all possible interleavings are considered. However, if we want to apply more aggressive POR techniques, we can add constraints to $SCH$ in order to force the removal of interleavings that do not contribute to checking a given property. In our running example, we can add the constraint $ts_{1}=2$ and $ts_{2}=1$ to remove the interleaving $\left\{Y_{1}\right\}$ (see Figure~\ref{figure:all-interleavings}(b)), which does not contribute to check the assertion in line 10 of Figure~\ref{figure:A-multi-threaded-C-program}.

\subsection{UW Approach}
\label{sec:UnderapproximationandWidening}

The core idea of the under-approximation and widening (UW) approach is to consider a series of under-approximations of a given model by encoding additional literals into the verification condition $\psi$ and by extracting the proof objects generated from an SMT solver~\cite{Z08}. We define $\psi'$ as an underapproximated model of $\psi$, i.e., $\psi' = \psi \wedge \bigwedge^{i=0}_{n} l_{i}$ where $l_{1}, l_{2}, \ldots, l_{n}$ are additional literals that guard the program statements of each thread. Similar to the schedule guards described in the previous section, these literals also control the symbolic execution: a program statement is executed only if the literal and its corresponding guard are enabled. Therefore, we can see that if $\psi$ is unsatisfiable, then $\psi'$ is also unsatisfiable, i.e., there is no assignment to the literals $l_{1}, l_{2}, \ldots, l_{n}$ that make $\psi'$ satisfiable. However, it is possible that $\psi$ is satisfiable while $\psi'$ is not, due to the additional literals. Thus, $\psi'$ can be thought of as an underapproximation of $\psi$ and each satisfying assignment of $\psi'$ is also a satisfying assignment to $\psi$. These additional literals then allow us to guide the widening process according to the variables that participate in the proof of unsatisfiability produced by the SMT solver. In the formal description, we rewrite formula (\ref{bounded-model-checking}) as

\begin{equation}
\label{bounded-model-checking-underapproximations}
\psi^{k} = \overbrace{I\left(s_{0}\right) \wedge \bigwedge^{n}_{j=1} \bigwedge^{k-1}_{i=0} \gamma'_{j} \left(s_{i},s_{i+1}\right)}^{constraints} \wedge \overbrace{\bigwedge^{n}_{j=1} P_{j}\left(s_{k}\right)}^{property} \wedge \overbrace{\bigwedge_{\forall i \in T} \bigwedge_{\forall j \in I} l_{ij}}^{UW model}
\end{equation}
\noindent where $l_{ij} \in L$ are literals that encode the program statements of each thread. We denote the set of threads by $T$, the set of program statements $S$, and the set of control literals by $L$. In the example of Figure~\ref{figure:control-point}, $T = \left\{T_{X}, T_{Y}\right\}$, $S = \left\{X_{0}, X_{1}, Y_{0}, Y_{1}\right\}$, and $L = \left\{ l_{X_{0}}, l_{Y_{0}}\right\}$. Note that the way that we encode the underapproximation differs from~\cite{GrumbergLST05}. The authors in~\cite{GrumbergLST05} encode an underapproximation using $m \times n$ control literals, where $m$ is the number of control points that guard each program statement and $n$ is the number of processes. In our encoding, we use the same guards as in the schedule recording approach as control literals, we use $l_{X_{0}} = \left(ts_{1} = 2\right) \: \&\& \: \left(ts_{2} = 1\right)$ and $l_{Y_{0}} = \left(ts_{1} = 2\right)$. We also use information from the front-end (as described in Section~\ref{sec:LazyApproach}) to reduce substantially the number of control literals required. If we were to include a control literal for each statement as in~\cite{GrumbergLST05}, then our solution might not scale in practice to large software systems.

The main difference between the schedule recording and the UW approaches is that schedule remains fixed and is by default set to true while the UW model is updated based on the information extracted from the proof and is initially set to false. The widening process then works as follows. Initially, each literal in $L$ is set to be false, because we aim to minimize the number of interleavings. At every state, we only consider the thread with the smallest index that has enabled transitions and only expand those. In our running example, we first execute program statement $Y_{1}$ because the tests $x > 2$ and $x > 3$ are false and consequently no other thread has enabled transitions. As the global variable $x$ is set initially to 2 (line 3), at the first step we consider that only program statement $Y_{1}$ is expanded from the initial state and build formula (\ref{bounded-model-checking-underapproximations}) by encoding $Y_{1}$ statement as $l_{Y_{0}} \Rightarrow \left(x=3\right)$. After that, we invoke the SMT solver to extract the unsatisfiable core and check that $l_{Y_{0}}$ participated in the proof of unsatisfiability. In the second iteration of our algorithm, we remove $l_{Y_{0}}$ from $L$ in order to continue to the next iteration so that $l_{Y_{0}}$ can now become either true or false (while the others must remain false). Afterwards, we execute symbolically program statements $X_{0}; X_{1}; Y_{0}$ and build formula (\ref{bounded-model-checking-underapproximations}). We check that $l_{X_{0}}$ participated in the proof of unsatisfiability. At the next iteration, we remove $l_{X_{0}}$ from $L$, execute symbolically program statements $Y_{0}; X_{0}; X_{1}$ and build formula (\ref{bounded-model-checking-underapproximations}). At this iteration, we have found a violation of the property and the UW procedure terminates; otherwise the procedure would continue until none of the additional literals in $L$ participate in the proof of unsatisfiability. It means that the procedure does not rely on the underapproximation itself and concludes that the property holds.

\subsection{Partial Order Reduction}
\label{sec:PartialOrderReduction}

In the modelling of multi-threaded software, we consider that any of the threads $j \in T$ is able to make a transition and then we have to compute all states for which a thread $j$ exists, (i.e., $\bigwedge^{n}_{j=1} \gamma_{j} \left(s_{i},s_{i+1}\right)$). The problem is that the number of states to be explored can grow dramatically with the number of program statements and threads. The purpose of the Partial-Order Reduction (POR) technique~\cite{Clarke00,CookKS05,Godefroid95,Peled93} is to reduce the number of states that have to be explored. This is done in a way that if the property holds on the reduced model, it also holds on the original model. 

In our SMT-based BMC framework, as threads communicate only through global variables, we apply partial order reduction (POR) techniques at two levels in our algorithm. At the first level, we apply the visible instruction analysis POR (VI-POR)~\cite{Peled93}, which removes the interleavings of instructions that do not affect the global variables, i.e., we remove transitions which are independent from transitions made by any other thread. An instruction is \textit{visible} if it accesses a global variable, and it is invisible otherwise. At the second level, we apply the read-write analysis POR (RW-POR)~\cite{CookKS05} in which two (or more) independent interleavings can be safely merged into one. In order to implement RW-POR, we compute the sets of variables written ($WR_{j}$) and read ($RD_{j}$) by each of the threads. If $WR_{j} \cap \left(\bigcup_{k \neq t} RD_{k} \cup WR_{k}\right) = \emptyset$ and $RD_{j} \cap \bigcup_{k \neq t} WR_{k} = \emptyset$, i.e., if the intersection between the set of visible variables that are written and read by thread $j$ and all other threads is empty, then we only explore the successors generated by executing $j$ and all other transitions can be safely ignored. 

There are six possible combinations of visible instructions of different threads, as shown in Table~\ref{table:Read-write-analysis-of-equivalence-of-interleavings}. There are three particular situations to consider when we generate the interleavings: \textit{(i)} two read operations will not modify the state, so they will always generate equivalent interleavings, \textit{(ii)} two program statements accessing different variables are independent w.r.t. their execution states, thus these two program statements always generate equivalent interleavings with both execution orders, \textit{(iii)} two instructions accessing same variable (i.e., with read-write and write-write relations) will generate non-equivalent interleavings. In these cases, the read-write relation actually causes read-write races and the write-write relation causes the write-write races. In summary, only two types of relations will generate non-equivalent interleavings, while all other four types of relations generate equivalent interleavings. Those redundant interleavings are simply removed in our approach.

\begin{table}[t!]
\setlength{\tabcolsep}{4pt}
\renewcommand\arraystretch{1.18}
\centering {\small
\begin{tabular}{|c|c|c|c|}
\hline
Access Relations & Read-read & Read-write & Write-write\\
\hline
Same variable & Equivalent & Non-equivalent & Non-equivalent \\
\hline
Different variables & Equivalent & Equivalent & Equivalent \\
\hline \end{tabular} }
\caption{Read-write analysis of interleavings equivalence between visible instructions.}
\label{table:Read-write-analysis-of-equivalence-of-interleavings}
\end{table}

Both PORs described above work best in conjunction with an alias analysis. However, at this point in our work, we do not have one implemented. We thus assume that the actual thread parameters are not aliased to global variables or to each other. In addition, we do not remove redundant interleavings originating from pointer aliasing.

%% file: 03-synchronization-primitives.tex
\section{Modelling Synchronization \\ Primitives in Pthread}
\label{sec:ModelingSynchronizationPrimitives}

This section presents our modelling of the synchronization primitives of the Pthread library~\cite{Mueller93alibrary}. We assume that the library function implementations are correct and focus our effort only on verifying client programs that use them. We thus provide an instrumented model of the Pthread functions and use this to model check the client code. We show, in our experiments, that our modelling is able to detect incorrect use of the functions and is also able to detect blocking operations that can lead to global deadlocks.

\subsection{Modelling Mutex Locking Operations}
\label{sec:ModellingMutexOperation}

The Pthread library supports two functions to implement mutual exclusion between threads, $pthread\_mutex\_lock$ and $pthread\_mutex\_unlock$. The argument to these functions is a C data structure called \textit{mutex} that, in our modelling, has two states, ``locked'' and ``unlocked''. The $pthread\_mutex\_lock$ locks the mutex if it is unlocked; otherwise it blocks the current thread until the mutex is released and can then be locked successfully again. The $pthread\_mutex\_unlock$ unlocks the mutex that was locked previously by the same thread.

Execution paths are considered to be blocked on a mutex when the thread tries to lock a mutex that has already been locked by other threads. Such blocking paths are also called non-wait-free paths. In order to model mutex operations, we apply the notion of wait-free paths as proposed initially in~\cite{RabinovitzG05}. However, in contrast to~\cite{RabinovitzG05}, our approach is able to model check multi-threaded programs that make use of mutexes, can handle more than two threads, can detect deadlocks, and does not require the user to run the model checker twice in order to detect different types of bugs (``regular'' and concurrency bugs).

\begin{figure}[ht]
\centering
\includegraphics[scale=0.22]{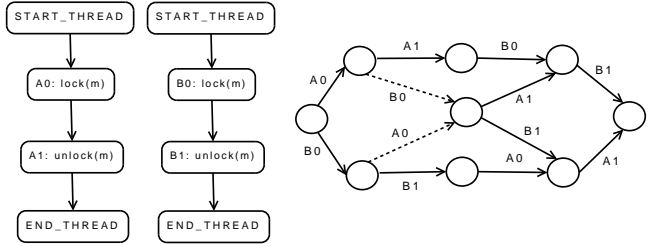}
\caption{Execution paths blocking on a mutex.}
\label{figure:Execution-paths-with-blocking-on-mutex}
\end{figure}

To explain how mutexes are encoded in our SMT-based BMC framework, we consider the example in Figure~\ref{figure:Execution-paths-with-blocking-on-mutex}. In this example, both threads $T_{A}$ and $T_{B}$ lock and unlock the same mutex $m$. The execution paths $A_{0}; A_{1}; B_{0}; B_{1}$ and $B_{0}; B_{1};$ $A_{0}; A_{1}$ are unblocked while the others are blocked paths. However, instead of blocking the execution paths starting with $A_{0}; B_{0}$ and $B_{0}; A_{0}$, we simple ignore the state of the mutex, so that we do not block the remaining instructions, and just lock it (again). In $pthread\_mutex\_unlock$, we simply check if the mutex is already locked and if so, we release the lock; otherwise, we have detected an error.

This modelling is sufficient to find bugs related to data races. However, it is not able to detect deadlocks. In order to detect global deadlock situations caused by the wrong use of the mutexes, we need to look in more detail at the possible states that a thread can be in with our modelling: \textit{(i)} \textit{Join state:} The thread is waiting for thread termination; \textit{(ii)} \textit{Lock state:} The thread is waiting for a mutex to be unlocked; \textit{(iii)} \textit{Wait state:} The thread is waiting for a signal or broadcast to wake up; \textit{(iv)} \textit{Exit state:} The thread has already exited; \textit{(v)} \textit{Free state:} The thread is not in any of the above four states and is free to execute its instructions. A thread is blocked if it is in one of the \textit{join}, \textit{lock} or \textit{wait} states, and is supposed to be running if it is not in \textit{exit} state. Global deadlock occurs if there is no running thread in the free state, i.e., the number of blocked threads is equal to the number of running threads. In order to model deadlock, counts of both blocked threads and running threads are maintained with global variables. Figure~\ref{figure:Modelling-lock-unlock-operations} presents our modelling of $pthread\_mutex\_lock$ to detect global deadlock with mutexes. We define $mutex\_lock\_field$ and $mutex\_count\_field$ as a C macro in lines 1 and 3 respectively.

\begin{figure}[ht]
\begin{lstlisting}
#define mutex_lock_field(a) 
					((a).__data.__lock)
#define mutex_count_field(a) 
					((a).__data.__count)
int pthread_mutex_lock(pthread_mutex_t *mutex)
{
  static _Bool unlocked = 1;
  extern unsigned int trds_in_run;
  if (!deadlock)
  {
	  atomic_begin();
	  unlocked = (mutex_lock_field(*mutex)==0);
	  if (unlocked)
		  mutex_lock_field(*mutex)=1;
		else
		  mutex_count_field(*mutex)++;
    atomic_end();

    atomic_begin();
    if (!unlocked)
    {
      if (!mutex_lock_field(*mutex))
        mutex_count_field(*mutex)--;        
      deadlock = (mutex_count_field(*mutex)
      						<trds_in_run);
      assert(deadlock);
    }
    atomic_end();
  }
  return 0;
}
int pthread_mutex_unlock(pthread_mutex_t *mutex)
{
  atomic_begin();
  assert(mutex_lock_field(*mutex));
  mutex_lock_field(*mutex)=0;
  atomic_end();
  return 0;
}

\end{lstlisting}
\caption{Modelling mutex lock and unlock operations to detect global deadlock.}
\label{figure:Modelling-lock-unlock-operations}
\end{figure}

We use the \textit{count} field of the $pthread\_mutex\_t$ data structure to count the number of threads that are in the lock state due to this mutex, and $trds\_in\_run$ to check the global number of threads that are currently running. Initially, the mutex is unlocked and we only lock it after the first call to $pthread\_mutex\_lock$. In subsequent calls, we increase the count field, allow context switches, check if the mutex was unlocked, and then assert $count$ $<$ $trds\_in\_run$. If the assertion fails, a global deadlock was detected (i.e., a thread is blocked by a lock operation on a mutex and the required mutex never gets unlocked by the thread that owns it, either because the locking thread has exited or because it has been blocked by another operation). If the assertion holds, we then eliminate this execution as described above. The modelling of the $pthread\_mutex\_unlock$, which is similar to~\cite{RabinovitzG05}, is shown at the bottom of Figure~\ref{figure:Modelling-lock-unlock-operations}. 

\subsection{Modelling Conditional Waiting}
\label{sec:ConditionalWaiting}

In the Pthread library, we consider functions from conditional waiting: $pthread\_cond\_wait$, $pthread\_cond\_signal$, and $pthread\_cond\_broadcast$. The arguments to the function $pthread\_cond\_wait$ are two data structures called \textit{cond} and \textit{mutex} where, in our modelling, \textit{cond} has also two states, ``locked'' and ``unlocked''. The others functions have only the argument \textit{cond}. Our modelling of the conditional waiting operation also employs the notion of wait-free execution paths. The function $pthread\_cond\_wait$ is used to block the thread on a condition variable and the blocked thread is awakened only if another thread calls signal or broadcast. If there are several threads that are blocked on a condition variable, then the $pthread\_cond\_signal$ call unblocks at least one of them (but there is no guarantee of which one will be woken up due to the scheduling policy) while the function $pthread$\_$cond$\_$broadcast$ call unblocks all threads currently blocked on the specified condition variable.

Figure~\ref{figure:Modelling-conditional-waiting-signal-operations} shows our modelling for the wait operation primitive. We consider that initially there is no deadlock (see line 4) and whenever a thread calls $pthread\_cond\_wait$, we atomically lock the condition variable \textit{cond}, assert that the \textit{mutex} is currently locked, and then release the \textit{mutex} so that other threads that access that \textit{mutex} can make progress (i.e., wait-free execution). Afterwards, we allow context switches and we then check whether the number of threads in wait state (i.e., threads that are waiting for a signal or broadcast to wake up) is less than the total number of the threads that are currently running.

\begin{figure}[ht]
\begin{lstlisting}
int pthread_cond_wait(pthread_cond_t *cond, 
											  pthread_mutex_t *mutex)
{
  static _Bool deadlock=0;
  extern unsigned int trds_in_run;    
  if (!deadlock)
  {
    atomic_begin();
    cond_lock_field(*cond)=1;
    assert(mutex_lock_field(*mutex));
    mutex_lock_field(*mutex)=0;
    cond_nwaiters_field(*cond)++;    
    atomic_end();
    atomic_begin();
    if (cond_lock_field(*cond))
    {
	    deadlock = (cond_nwaiters_field(*cond) <
	    						 trds_in_run);
      assert(deadlock);
    }
    assume(deadlock && 
    				cond_lock_field(*cond)==0);
    atomic_end();
    mutex_lock_field(*mutex)=1;
  }
}
int pthread_cond_signal(pthread_cond_t *cond)
{
  atomic_begin();
	cond_lock_field(*cond)=0;
	cond_nwaiters_field(*cond)--;
	atomic_end();
  return 0;
}

\end{lstlisting}
\caption{Modelling conditional waiting and signal operations to detect global deadlock.}
\label{figure:Modelling-conditional-waiting-signal-operations}
\end{figure}

In order to model signal operations, we simply release the condition variable and decrement the number of threads that were locked due to the specified condition variable. The modelling of the conditional signal operation is shown in Figure~\ref{figure:Modelling-conditional-waiting-signal-operations} as well.

In order to model broadcast operations, we create an additional global variable called $broadcast\_id$, which records the number of broadcast operations that have executed and also gets incremented inside the function $pthread\_cond\_broadcast$. In the wait operation the thread firstly records the current $broadcast\_id$ and is then forced to make context switches to other threads. When the context is switched back to the current thread, an assertion checks if a broadcast operation has occurred by checking whether the current value of variable $broadcast\_id$ is greater than the recorded $broadcast\_id$. The deadlock is detected if there is no execution path with broadcast operations.

%% file: 04-experimental-evaluation.tex
\section{Experimental Evaluation}
\label{05-experimental-results}

We have implemented the lazy, schedule recording, and UW approaches described in Section~\ref{03-proposed-method} in our ESBMC\footnote{Available at http://users.ecs.soton.ac.uk/lcc08r/esbmc/} (Efficient SMT-Based Bounded Model Checker) tool that supports the SMT logics QF$\_$AUFBV as well as QF$\_$AUFLIRA from the SMT-LIB~\cite{smtlib09}. In our experiments, we have used ESBMC v1.3 together with the SMT solver Z3~\cite{Z08}. 


The experimental evaluation of our work consists of two parts. In Section~\ref{sec:ComparisonToMPORAndPPOR}, we compare our approaches against the Monotonic Partial Order Reduction (MPOR)~\cite{KahlonWG09} and Peephole Partial Order Reduction (PPOR) \cite{WangYKG08} that are implemented in a SMT-based bounded model checker using the Yices SMT solver~\cite{Yices06}. In Section~\ref{sec:ComparisonToSATABS}, we compare our approaches against SATABS version 2.4~\cite{cksy2005} connected to Cadence SMV \cite{McMillan10}, which is a state-of-the-art C model checker and supports the verification of multi-threaded software with shared variables using the CEGAR technique. All experiments were conducted on an otherwise idle Intel Xeon 5160, 3GHz server with 4 GB of RAM running Linux OS. For all benchmarks, the time limit has been set to 3600 seconds for each individual property. All times given are wall clock time in seconds as measured by the unix \textit{time} command through a single execution.

\subsection{Comparison to MPOR and PPOR}
\label{sec:ComparisonToMPORAndPPOR}

We use the dining philosophers model to evaluate our approaches against MPOR and PPOR. Since the benchmarks used in~\cite{KahlonWG09} are not available, we re-implemented them as described there. An implementation is available at \texttt{users.ecs.soton.ac.uk/lcc08r/esbmc}. Each philosopher has its own local variables, and they communicate only through a global shared array of forks. This version guarantees the absence of deadlocks. As in~\cite{KahlonWG09}, we also check two properties: \textit{(i)} whether all philosophers can eat simultaneously (this property does not hold, i.e., the verification condition is unsatisfiable) and \textit{(ii)} whether all philosophers have eaten at least once (this property holds, i.e., the verification condition is satisfiable). The authors in~\cite{KahlonWG09} run their experiments on a workstation with 2.8 GHz Xeon processor and 4GB of RAM memory running Linux OS. In order to make the results comparable, we scale their times in Table~\ref{table:results-of-the-benchmark-part}. We give both original (in brackets) and scaled timings.

Table~\ref{table:results-of-the-benchmark-part} shows the detailed results of the comparison between MPOR, PPOR, and the three ESBMC approaches. The first column $\#$L gives the number of lines of code, while the second column $\#$T reports the total number of threads. The \textit{Time} column provides the time in seconds while the column $\#$I provides the total number of generated interleavings and the column $\#$IF the total number of failed interleavings. The column \textit{Iter} gives the number of iterations to prove or disprove the property in the UW approach.

\begin{table*}[t]
\renewcommand\arraystretch{1.18}
\setlength{\tabcolsep}{4pt}
\begin{center} {\small
\begin{tabular}{|c|l|c|c||c||c||c|c||c||c|c|}
\hline
  & & & & \multicolumn{1}{c||}{MPOR} & \multicolumn{1}{c||}{PPOR} & \multicolumn{2}{c||}{Lazy} & \multicolumn{1}{c||}{Schedule} & \multicolumn{2}{c|}{UW} \\ \cline{5-11}
  & Module & $\#$L & $\#$T & Time & Time & Time & \multicolumn{1}{c||}{$\#$I/$\#$IF} & Time & Time & Iter \\ \cline{1-11} 
1 & din$\_$phil2$\_$unsat & 63 & 2 & 0.2 (0.2) & \textbf{0.1} (0.1) & 0.2 & 2/0 & 0.2 & 0.2 & 1 \\
\hline
2 & din$\_$phil3$\_$unsat & 63 & 3 & 0.8 (0.9) & 1.0 (1.1) & 0.5 & 6/0 & \textbf{0.4} & 0.5 & 1 \\
\hline
3 & din$\_$phil4$\_$unsat & 63 & 4 & 5.0 (5.3) & 41.9 (44.9) & 2 & 24/0 & \textbf{1.6} & \textbf{1.6} & 1 \\
\hline
4 & din$\_$phil5$\_$unsat & 63 & 5 & 21.4 (22.9) & 138.7 (148.6) & 11.1 & 120/0 & \textbf{8.3} & 8.7 & 1 \\
\hline
5 & din$\_$phil6$\_$unsat & 63 & 6 & \textbf{48.8} (52.3) & 470.4 (504.4) & 74 & 720/0 & 115.8 & 115.4 & 1 \\
\hline
6 & din$\_$phil7$\_$unsat & 63 & 7 & \textbf{150.8} (161.6) & TO & 574.1 & 5040/0 & TO & TO & 0 \\
\hline
7 & din$\_$phil2$\_$sat & 63 & 2 & 0.1 (0.1) & 0.1 (0.1) & \textbf{0.2} & 2/2 & 0.2 & 0.4 & 3 \\
\hline
8 & din$\_$phil3$\_$sat & 63 & 3 & 1.2 (1.3) & 0.3 (0.3) & \textbf{0.2} & 6/6 & 0.5 & 1.6 & 4 \\
\hline
9 & din$\_$phil4$\_$sat & 63 & 4 & 8.9 (9.5) & 3.6 (3.8) & \textbf{0.2} & 24/24 & 1.8 & 2.6 & 5 \\
\hline
10 & din$\_$phil5$\_$sat & 63 & 5 & 88.4 (94.7) & 57.6 (61.7) & \textbf{0.3} & 120/120 & 8.8 & 33 & 6 \\
\hline
11 & din$\_$phil6$\_$sat & 63 & 6 & 294.4 (315.4) & 2130.8 (2283) & \textbf{0.3} & 720/720 & 105.6 & 105.2 & 1 \\
\hline
12 & din$\_$phil7$\_$sat & 63 & 7 & 1136.8 (1218) & TO & \textbf{0.3} & 5040/5040 & TO & TO & 0 \\
\hline
\end{tabular} }
\end{center}
\caption{Results of the comparison between MPOR, PPOR, lazy, schedule, and UW ESBMC}
\label{table:results-of-the-benchmark-part}
\end{table*}

As we can see in Table~\ref{table:results-of-the-benchmark-part}, our approaches perform equivalently to MPOR to check the first property of the model until we set the number of philosophers to 5. If we continue increasing the number of philosophers,  MPOR performs better than our approaches. However, our three approaches perform better than PPOR to check the first property. In addition, our lazy ESBMC scales significantly better than the other approaches to check the second property of the dining philosophers model, i.e., whether all philosophers have eaten at least once. We also show in column $\#$I/$\#$IF that all interleavings generated by our lazy ESBMC are satisfiable. Our UW and schedule ESBMC also performs better than MPOR and PPOR until we set the number of philosophers to 6. In summary, our lazy approach outperforms both MPOR and PPOR for those benchmarks that generate satisfiable formulae and is still comparable to MPOR and PPOR when the generated formulae are unsatisfiable.

\subsection{Comparison to SATABS}
\label{sec:ComparisonToSATABS}

In order to evaluate our approaches against SATABS, we used a number of multi-threaded programs taken from standard benchmark suites (see Table~\ref{table:results-of-the-benchmark-part-ii}). Programs 1-12 are an implementation of the dining philosophers as described in Section~\ref{sec:ComparisonToMPORAndPPOR}. In the dining philosophers implementation, we set the number of philosophers (threads) to $2, 3, \ldots, 7$ and compare the runtime performance of the three approaches against SATABS. The programs 13-22 are taken from the benchmark suite of the INSPECT tool~\cite{Inspect10}. This suite contains programs with two or more threads as well as mutex and condition synchronization primitives from the Pthread library. The programs 23 and 24 are taken from the Helgrind benchmark suite~\cite{Helgrind10} and they contain concurrency bugs related to lock and unlock operations. It is important to note that most of these benchmarks contain data dependencies among the threads (i.e., the threads access the global variables).

Table~\ref{table:results-of-the-benchmark-part-ii} shows the detailed results of the comparison between UW, Lazy, and schedule ESBMC as well as SATABS. We do not run the programs 19-22 with SATABS because it does not support the condition synchronization primitive. It is also important to point out that the verification times of the programs 1-12 in Table~\ref{table:results-of-the-benchmark-part-ii} differ from Table~\ref{table:results-of-the-benchmark-part} because instead of checking a single property, here we check properties related to mutex operations and array bounds, which can be automatically generated by both tools, SATABS and ESBMC. Hence, the column $\#$P gives the number of properties to be verified for each multi-threaded C program. The \textit{Time} column provides the time in seconds to check all properties of a given program and \textit{Failed} indicates how many properties failed during the verification process. Here, properties can fail for two reasons: either due to a time out (TO) or due to memory out (MO). 

\begin{table*}[t]
\renewcommand\arraystretch{1.18}
\setlength{\tabcolsep}{4pt}
\begin{center} {\small
\begin{tabular}{|c|l|c|c||r|r|r|r||r|r|r|r|r||r|r|r|r||r|r|r|r|r|}
\hline
 & & & & \multicolumn{4}{c||}{SATABS} & \multicolumn{5}{c||}{UW} & \multicolumn{4}{c||}{Schedule} & \multicolumn{5}{c|}{Lazy}\\  \cline{5-22}
 & & & & Time & \multicolumn{3}{c||}{$\#$P} & Time & \multicolumn{3}{c|}{$\#$P} & & Time & \multicolumn{3}{c||}{$\#$P} & Time & \multicolumn{3}{c|}{$\#$P} & \multicolumn{1}{c|}{$\#$I} \\ \cline{5-22}
 & Module & $\#$L & $\#$T & \multicolumn{1}{c|}{\rotatebox{90}{Total}} & \multicolumn{1}{c|}{\rotatebox{90}{Passed}} & \multicolumn{1}{c|}{\rotatebox{90}{Violated}} & \multicolumn{1}{c||}{\rotatebox{90}{Failed}} & \multicolumn{1}{c|}{\rotatebox{90}{Total}} & \multicolumn{1}{c|}{\rotatebox{90}{Passed}} & \multicolumn{1}{c|}{\rotatebox{90}{Violated}} & \multicolumn{1}{c|}{\rotatebox{90}{Failed}} & \multicolumn{1}{c||}{\rotatebox{90}{Iter}} & \multicolumn{1}{c|}{\rotatebox{90}{Total}} & \multicolumn{1}{c|}{\rotatebox{90}{Passed}} & \multicolumn{1}{c|}{\rotatebox{90}{Violated}} & \multicolumn{1}{c||}{\rotatebox{90}{Failed}} & \multicolumn{1}{c|}{\rotatebox{90}{Total}} & \multicolumn{1}{c|}{\rotatebox{90}{Passed}} & \multicolumn{1}{c|}{\rotatebox{90}{Violated}} & \multicolumn{1}{c|}{\rotatebox{90}{Failed}} & \multicolumn{1}{c|}{\rotatebox{90}{$\#$I / $\#$IF}} \\\hline
1 & din$\_$phil2$\_$unsat & 63 & 2 & 26 & 13 & 0 & 0 & 0.3 & 28 & 0 & 0 & 1 & \textbf{0.2} & 28 & 0 & 0 & 0.3 & 28 & 0 & 0 & 2/0\\
\hline 
2 & din$\_$phil3$\_$unsat & 63 & 3 & 26 & 13 & 0 & 0 & 0.6 & 28 & 0 & 0 & 1 & \textbf{0.5} & 28 & 0 & 0 & 0.7 & 28 & 0 & 0 & 6/0\\
\hline 
3 & din$\_$phil4$\_$unsat & 63 & 4 & 26 & 13 & 0 & 0 & 2.3 & 28 & 0 & 0 & 1 & \textbf{1.7} & 28 & 0 & 0 & 2.7 & 28 & 0 & 0 & 24/0\\
\hline 
4 & din$\_$phil5$\_$unsat & 63 & 5 & 26.2 & 13 & 0 & 0 & 12.3 & 28 & 0 & 0 & 1 & \textbf{9.4} & 28 & 0 & 0 & 14 & 28 & 0 & 0 & 120/0\\
\hline 
5 & din$\_$phil6$\_$unsat & 63 & 6 & \textbf{26.4} & 13 & 0 & 0 & 142 & 28 & 0 & 0 & 1 & 123.8 & 28 & 0 & 0 & 91.7 & 28 & 0 & 0 & 720/0\\
\hline
6 & din$\_$phil7$\_$unsat & 63 & 7 & \textbf{27.1} & 13 & 0 & 0 & TO & 0 & 0 & 28 & 0 & MO & 0 & 0 & 28 & 719.2 & 28 & 0 & 0 & 5040/0\\
\hline
7 & din$\_$phil2$\_$sat & 63 & 2 & 30.1 & 18 & 0 & 0 & 0.5 & 28 & 1 & 0 & 3 & 0.3 & 28 & 1 & 0 & \textbf{0.2} & 28 & 1 & 0 & 2/2\\
\hline 
8 & din$\_$phil3$\_$sat & 63 & 3 & 28.4 & 18 & 0 & 0 & 2.2 & 28 & 1 & 0 & 4 & 0.7 & 28 & 1 & 0 & \textbf{0.2} & 28 & 1 & 0 & 6/6\\
\hline 
9 & din$\_$phil4$\_$sat & 63 & 4 & 28.4 & 18 & 0 & 0 & 3.05 & 28 & 1 & 0 & 5 & 2.4 & 28 & 1 & 0 & \textbf{0.2} & 28 & 1 & 0 & 24/24\\
\hline 
10 & din$\_$phil5$\_$sat & 63 & 5 & 28.6 & 18 & 0 & 0 & 42.5 & 28 & 1 & 0 & 6 & 14.4 & 28 & 1 & 0 & \textbf{0.3} & 28 & 1 & 0 & 120/120\\
\hline 
11 & din$\_$phil6$\_$sat & 63 & 6 & 27.8 & 18 & 0 & 0 & 180 & 28 & 1 & 0 & 1 & 177.5 & 28 & 1 & 0 & \textbf{0.3} & 28 & 1 & 0 & 720/720\\
\hline 
12 & din$\_$phil7$\_$sat & 63 & 7 & 28.9 & 18 & 0 & 0 & MO & 28 & 1 & 0 & 0 & MO & 28 & 1 & 0 & \textbf{0.3} & 28 & 1 & 0 & 5040/5040\\
\hline\hline 
13 & carter01$\_$ok & 58 & 2 & 25 & 2 & 0 & 0 & \textbf{0.3} & 4 & 0 & 0 & 1 & \textbf{0.3} & 4 & 0 & 0 & \textbf{0.3} & 4 & 0 & 0 & 2/0\\
\hline
14 & lazy01$\_$ok & 48 & 3 & 3 & 1 & 0 & 0 & \textbf{0.2} & 4 & 0 & 0 & 1 & \textbf{0.2} & 4 & 0 & 0 & 0.3 & 4 & 0 & 0 & 6/0\\
\hline 
15 & phase01$\_$ok & 34 & 1 & 23 & 0 & 1 & 0 & \textbf{0.2} & 4 & 0 & 0 & 1 & \textbf{0.2} & 4 & 0 & 0 & \textbf{0.2} & 4 & 0 & 0 & 2/0\\
\hline 
16 & stateful01$\_$ok & 47 & 2 & 2 & 1 & 0 & 0 & \textbf{0.2} & 4 & 0 & 0 & 1 & \textbf{0.2} & 4 & 0 & 0 & \textbf{0.2} & 4 & 0 & 0 & 2/0\\
\hline 
17 & stateful06$\_$ok & 59 & 2 & TO & 0 & 0 & 2 & \textbf{0.2} & 5 & 0 & 0 & 1 & \textbf{0.2} & 5 & 0 & 0 & \textbf{0.2} & 5 & 0 & 0 & 2/0\\
\hline
18 & stateful20$\_$ok & 61 & 3 & TO & 0 & 0 & 2 & 5.7 & 5 & 0 & 0 & 1 & 3.3 & 5 & 0 & 0 & \textbf{0.7} & 5 & 0 & 0 & 6/6\\
\hline
19 & sync01$\_$ok & 62 & 2 & - & - & - & - & \textbf{0.2} & 4 & 0 & 0 & 1 & 0.3 & 4 & 0 & 0 & 0.3 & 4 & 0 & 0 & 2/0\\
\hline
20 & sync01$\_$bad & 62 & 2 & - & - & - & - & \textbf{0.3} & 3 & 1 & 0 & 1 & \textbf{0.3} & 3 & 1 & 0 & \textbf{0.3} & 3 & 1 & 0 & 2/1\\
\hline
21 & sync02$\_$ok & 73 & 2 & - & - & - & - & \textbf{0.3} & 4 & 0 & 0 & 1 & \textbf{0.3} & 4 & 0 & 0 & \textbf{0.3} & 4 & 0 & 0 & 2/0\\
\hline
22 & sync02$\_$bad & 73 & 2 & - & - & - & - & \textbf{0.3} & 3 & 1 & 0 & 1 & \textbf{0.3} & 3 & 1 & 0 & \textbf{0.3} & 3 & 1 & 0 & 2/1\\
\hline\hline 
23 & tc10$\_$rec$\_$lock$\_$bad & 49 & 1 & \textbf{0.3} & 7 & 0 & 0 & \textbf{0.3} & 8 & 2 & 0 & 1 & \textbf{0.3} & 8 & 2 & 0 & \textbf{0.3} & 8 & 2 & 0 & 1/1 \\
\hline 
24 & tc14$\_$laog$\_$bad & 45 & 5 & 20 & 17 & 0 & 0 & TO & 0 & 0 & 28 & 2 & TO & 0 & 0 & 28 & \textbf{0.8} & 10 & 18 & 0 & 120/120\\
\hline 
\end{tabular} }
\end{center}
\caption{Results of the comparison between SATABS, UW, schedule, and lazy ESBMC}
\label{table:results-of-the-benchmark-part-ii}
\end{table*}

As we can see in Table~\ref{table:results-of-the-benchmark-part-ii}, our lazy ESBMC approach performs significantly better than the other approaches on benchmarks that contain bugs (i.e., the formula sent to the SMT solver is satisfiable). However, if there is no bug in the benchmark, then our schedule ESBMC approach performs better than the UW and lazy ESBMC, but not as good as SATABS for the dining philosophers benchmark. This indicates that our SMT-based BMC procedures do not scale well for problems of increasing complexity, i.e., for a large number of threads and data dependencies among the threads. However, SATABS times out for programs 17 and 18, and provides false results for programs 7-12, 15, 23, and 24, of which the last two contain deadlocks due to the incorrect use of lock and unlock operations. Based on that, we conclude that SATABS does not seem to explore all interleavings and also does not add additional checks for detecting deadlocks, which explains the better scaling for the dining philosophers benchmark.

We can see that our UW ESBMC algorithm outperforms SATABS in most of the multi-threaded programs from Table~\ref{table:results-of-the-benchmark-part-ii}, except for the programs 5, 6, 10, 11, and 12. However, in these programs SATABS provides false results as discussed above. In any case, it is important to note that when we enabled the proof generation feature of the SMT solver to extract the unsatisfiable cores, we always observed memory overhead and corresponding slowdowns, as also reported previously in~\cite{MouraB08}. Additionally, we observed that the performance of the UW ESBMC procedure can be significantly improved if we use heuristics to update the set of additional literals in $L$ to be used at the next iteration of the algorithm. However, at this point in time, we do not investigate further alternative ways of updating the set L. We set the maximum size of the unsatisfiable core to contain 500 control literals since the SMT solver Z3 fails with a segmentation fault when there are thousands of literals. This situation occurs only with the dining philosophers model when we set the number of philosophers to 6 or more. We reported this bug to the Z3 developers and they were already aware of this problem.

%% file: 05-related-work.tex
\section{Related Work}
\label{05-related-work}

SMT-based BMC is gaining popularity in the formal verification community due to the advent of sophisticated SMT solvers built over efficient SAT solvers~\cite{Z08}.\ Ganai and Gupta describe a verification framework for BMC which extracts high-level design information from an extended finite state machine (EFSM) and apply several techniques to simplify the BMC problem~\cite{Ganai06}.\ However, the authors use only the theory of integer and real arithmetic, which does not reflect precisely the ANSI-C semantics.\ Armando et al.\ also propose a BMC approach using SMT solvers for ANSI-C programs~\cite{Armando09}, but they only make use of linear arithmetic, arrays, records and restricted bit-vectors arithmetic and, as a consequence, their SMT-CBMC prototype does not address important constructs of the ANSI-C language.

Qadeer and Rehof present a pragmatic method to discover bugs in concurrent software in which the program analysis is restricted to executions with a bounded number of context switches~\cite{QadeerR05}.\ However, this method is incomplete since it considers the verification up to a given fixed context bound.\ In addition, the authors do not apply it to realistic and large concurrent software benchmarks and the integration of this context-bounded model checking algorithm into the explicit state model checker ZING~\cite{AndrewsQRX04} is left for future work.\ Rabinovitz and Grumberg describe an extension of CBMC to concurrent C programs~\cite{RabinovitzG05}, which translates C threads into SSA form and adds constraints for a bounded number of context-switches, as described in~\cite{AndrewsQRX04}.\ This approach, however, is limited to two threads and it requires additional constraints to bound the number of context switches and allowed interleavings into the formula to be sent to a SAT solver.

Ganai and Gupta describe a lazy method for modelling multi-threaded concurrent systems using shared variables~\cite{GanaiG08}, but this method is restricted to two threads.\ Gupta et al.~\cite{KahlonWG09} extend~\cite{GanaiG08,KahlonSG09} by supporting more than two threads and by combining dynamic partial order reduction with symbolic state space exploration.\ However, this method is incomplete since it considers the concurrency semantics up to the bounded depth as in~\cite{AndrewsQRX04,RabinovitzG05}.\ Grumberg et~al.\ propose an algorithmic method based on SAT and BMC to model check a multi-process system based on a series of under-approximated models~\cite{GrumbergLST05}.\ This approach, however, does not integrate partial order reduction algorithms to reduce redundant interleavings and it does not address the problem of model checking real-world embedded software in multi-core environments.

To the best of our knowledge, there is no work that considers a comprehensive SMT-based BMC formulation to verify multi-threaded software using a set of under-approximations and widening models as well as the integration of partial order reduction algorithms into the UW framework.\ In contrast to~\cite{GanaiG08,RabinovitzG05}, our method can handle more than two threads and can detect deadlock caused by the mutexes and conditions operations. Our main contribution is an algorithmic method and corresponding tools to verify multi-threaded software using SMT in order to combat the verification complexity.\

%% file: 06-conclusion.tex
\section{Conclusions and Future Work}
\label{06-conclusion}

Despite the large body of (theoretical) research in the verification of concurrent systems, there are only few tools that analyze multi-threaded programs with shared variables. In this work, we presented an extension of the ESBMC model checker to support the verification of multi-threaded software with shared variables, mutexes and conditions using an SMT-based BMC framework. We also described three approaches UW, lazy and eager SMT-based BMC implemented with partial-order reduction methods in which the final formula is well suited for using with the SMT solvers. Our experimental results show that our UW ESBMC approach outperforms the CEGAR approach implemented in the SATABS model checker. With the addition of deadlock detection in our modelling, we can find bugs that other previous approaches are not able to find. Moreover, our lazy ESBMC, which adds concurrency constraints lazily and incrementally, is able to find bugs quickly in non-trivial benchmarks. In future, we would like to explore in more depth the partial-order reduction methods, configure ESBMC for compatibility with any given compiler to break statements with multiple global variables, and investigate heuristics to update the set of additional literals in our UW ESBMC algorithm.